\documentclass[11pt,twoside]{article}
\usepackage{asp2010}

\resetcounters

\begin{document}

\title{The M4 Transition: Toward a comprehensive understanding of the
transition into the fully convective regime}
\author{
Keivan G.\ Stassun$^{1}$, Leslie Hebb$^1$, Kevin Covey$^2$, Andrew A.\ West$^3$,
Jonathan Irwin$^4$, Richard Jackson$^5$, Moira Jardine$^6$, Julien Morin$^7$,
Dermott Mullan$^8$, Neill Reid$^{9}$
\affil{$^1$Department of Physics \& Astronomy, Vanderbilt University,
VU Station B 1807, Nashville, TN  37235, USA}
\affil{$^2$Department of Astronomy, Cornell University, 226 Space Sciences 
Building, Ithaca, NY 14853, USA} 
\affil{$^3$Department of Astronomy, Boston University, 725 Commonwealth 
Ave, Boston, MA 02215, USA} 
\affil{$^4$Harvard-Smithsonian Center for Astrophysics, 60 Garden St., 
Cambridge, MA 02138, USA} 
\affil{$^5$Astrophysics Group, Keele University, Keele, Staffordshire ST5 5BG,
UK} 
\affil{$^6$School of Physics and Astronomy, University of St Andrews, 
St Andrews, Scotland KY16 9SS, UK} 
\affil{$^7$Dublin Institute for Advanced Studies, 
31 Fitzwilliam Place, Dublin 2, Ireland} 
\affil{$^8$Department of Physics and Astronomy, University of Delaware, 
Newark, DE 19716, USA} 
\affil{$^9$Space Telescope Science Institute, Baltimore, MD 21218, USA}
} 

\begin{abstract}
The difference in stellar structure above and below spectral type $\sim$M4
is expected to be a very important one, connected directly or
indirectly to a variety of observational phenomena in cool stars---such as
rotation, activity, magnetic field generation and topology, timescales
for evolution of these, and even the basic mass-radius relationship.
In this Cool Stars XVI Splinter Session, 
we aimed to use the M4 transition as an
opportunity for discussion about the interiors of low-mass stars and
the mechanisms which determine their fundamental properties.
By the conclusion of the session, several key points were elucidated.
Although M dwarfs exhibit significant
changes across the fully convective boundary, this ``M4 transition" is
not observationally sharp or discrete.
Instead, the properties of M dwarfs (radius, effective
temperature, rotation, activity lifetime, magnetic field strength and topology)
show smooth changes across M3--M6 spectral types.  
In addition, a wide range of stellar masses share
similar spectral types around the fully convective transition. 
There appears to be a second transition at M6--M8 spectral types, below which
there exists a clear dichotomy of magnetic field topologies.  Finally, 
we used the information and ideas presented in the session to construct
a framework for how the structure of an M~dwarf star, 
born with specific mass and chemical composition, responds to the
presence of its magnetic field, itself driven by a feedback process 
that links the star's rotation, interior structure, and field topology.

\end{abstract}

\section{Introduction}
The interior structure of a star is a function of its mass. Above 
a threshold mass of $\sim$0.3 M$_\odot$, the interior is expected to resemble 
that of the Sun, having a radiative zone and a convective envelope. In 
contrast, below the threshold mass the convection zone extends all the way 
to the core. 

Despite not having a
strong radiative/convective interface (like that found in the Sun),
mid- to late-type M dwarfs (spectral type of M4 and greater) are observed to 
have strong magnetic fields and subsequent surface heating that results in
observed magnetic activity (traced by strong line or continuum emission).
Recent theoretical modeling of stars with convective envelopes that extend
to the core (and a lack of a strong radiative/convective interface) were
able to produce strong, long lived, large scale magnetic fields \citep{Browning2008}.
Observational results also suggest that in mid- to late-type M dwarfs,
the fraction of the magnetic field in large scale components is larger
than in early type M dwarfs \citep{Donati2008,Reiners2009}.

These are just some examples of the phenomena and questions that play 
out around the M4 spectral type. 
The purpose of this Cool Stars XVI Splinter Session
was to use this specific transition point---the
M4 transition---as an opportunity for focused discussion and ``out of the box"
collective thinking about the interiors of low-mass stars, what we know,
what we don't know, what we wish we understood better. We intended this to
be a rich, fun session to bring together observational evidence from
multiple directions with the goal of disentangling the effects of multiple
physical phenomena and leading toward a clearer set of 
guiding questions for future work.

\section{The M4 transition}


Neill Reid began by reminding us of some of the basic properties of M 
dwarfs and of the basic predictions of theoretical stellar evolution models 
for M dwarf interior structure.
Most theoretical stellar evolution models predict the transition into the
fully convective regime to occur at a stellar mass of $\sim$0.3 M$_\odot$,
but differ in detail. For example, \citet{Chabrier1997} predict the
boundary to occur at $M \approx 0.35$ M$_\odot$ (spectral type M2), whereas
\citet{Dorman1989} place the fully convective boundary at 
$M \approx 0.25$ M$_\odot$ (spectral type M4). 
For the fiducial properties of M2 stars, refer to the benchmark
systems Gliese 22A \citep{Henry1993} and Gliese 411 \citep{Lane2001},
both with empirical masses of $\approx$0.36 M$_\odot$ and the latter with
an interferometrically measured radius. For the fiducial properties
of M4 stars, refer to the benchmark eclipsing binary CM Dra, comprising
two M4.5 stars with masses of 0.22 M$_\odot$ \citep[e.g.][]{Morales2009}.
Table~\ref{table:m4properties} gives
a summary of the physical properties of spectral type M2 and M4.5 stars.


\begin{table}[ht]
\caption{\label{table:m4properties}
Basic physical properties of exemplar M2 and M4.5 dwarfs.}
\smallskip
\begin{center}
{\small
\begin{tabular}{lccccccc}
\tableline
\noalign{\smallskip}
Name & Mass (M$_\odot$) & Radius (R$_\odot$) & $T_{\rm eff}$ & M$_V$ & 
$V-K$ \\
\noalign{\smallskip}
\tableline
\noalign{\smallskip}
GJ 22 A  & $ 0.352\pm0.036 $   & ...               & $3380\pm150$ & $10.99\pm0.06$ & 4.55  \\
GJ 411   & $ 0.403\pm0.020 $   & $0.393\pm0.008$   & $3828\pm100$ & $8.08\pm0.09 $ & 4.25 \\   
CM Dra A & $ 0.2310\pm0.0009 $ & $0.2534\pm0.0019$ & $3130\pm70$  & $12.78\pm0.05$ & 5.10 \\
CM Dra B & $ 0.2141\pm0.0010 $ & $0.2396\pm0.0015$ & $3120\pm70$  & $12.91\pm0.05$ & 5.10 \\
\noalign{\smallskip}
\tableline
\end{tabular}
}
\end{center}
\end{table}

\subsection{Basic stellar properties}
\begin{figure}[ht]
\plottwo{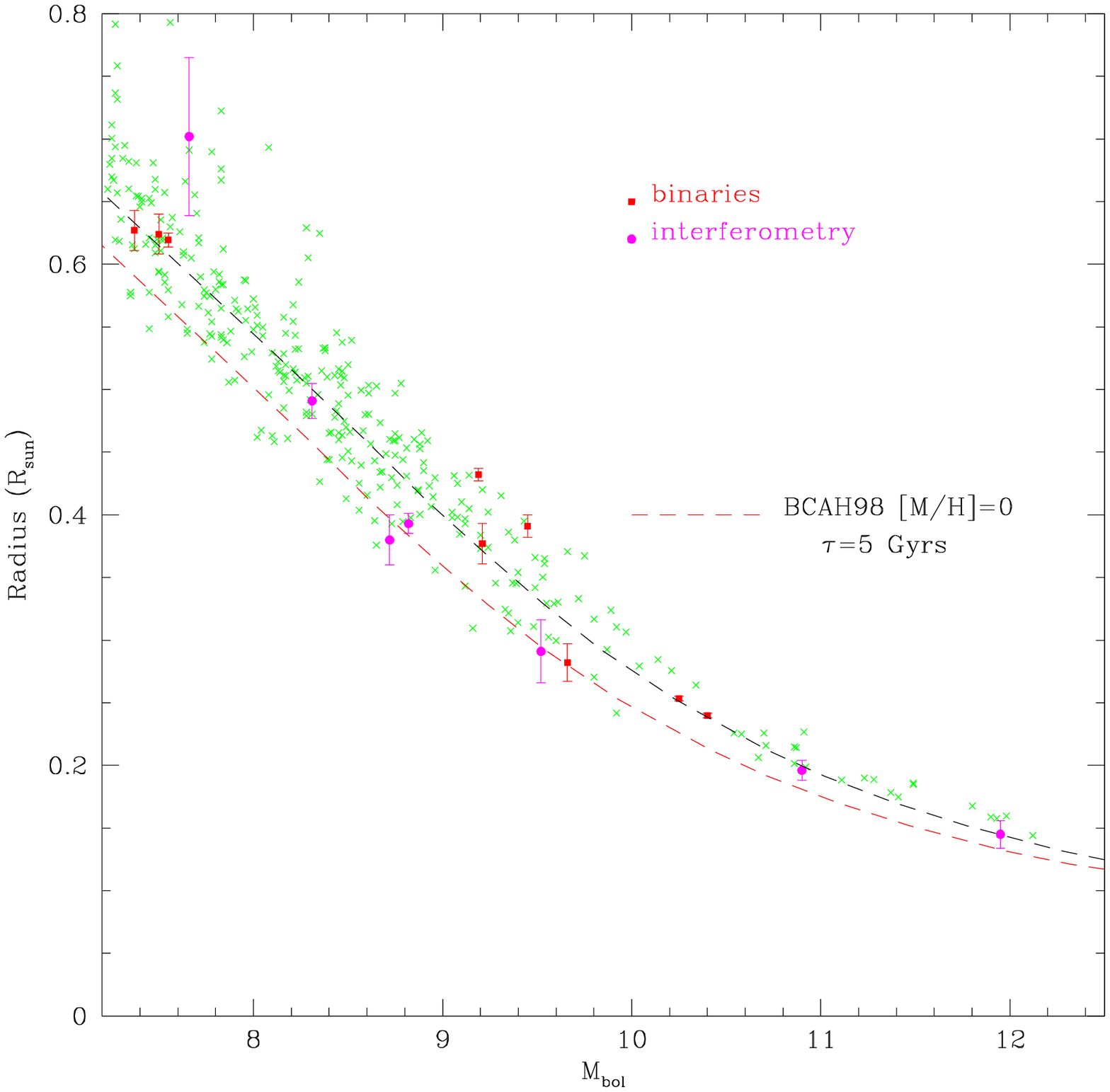}{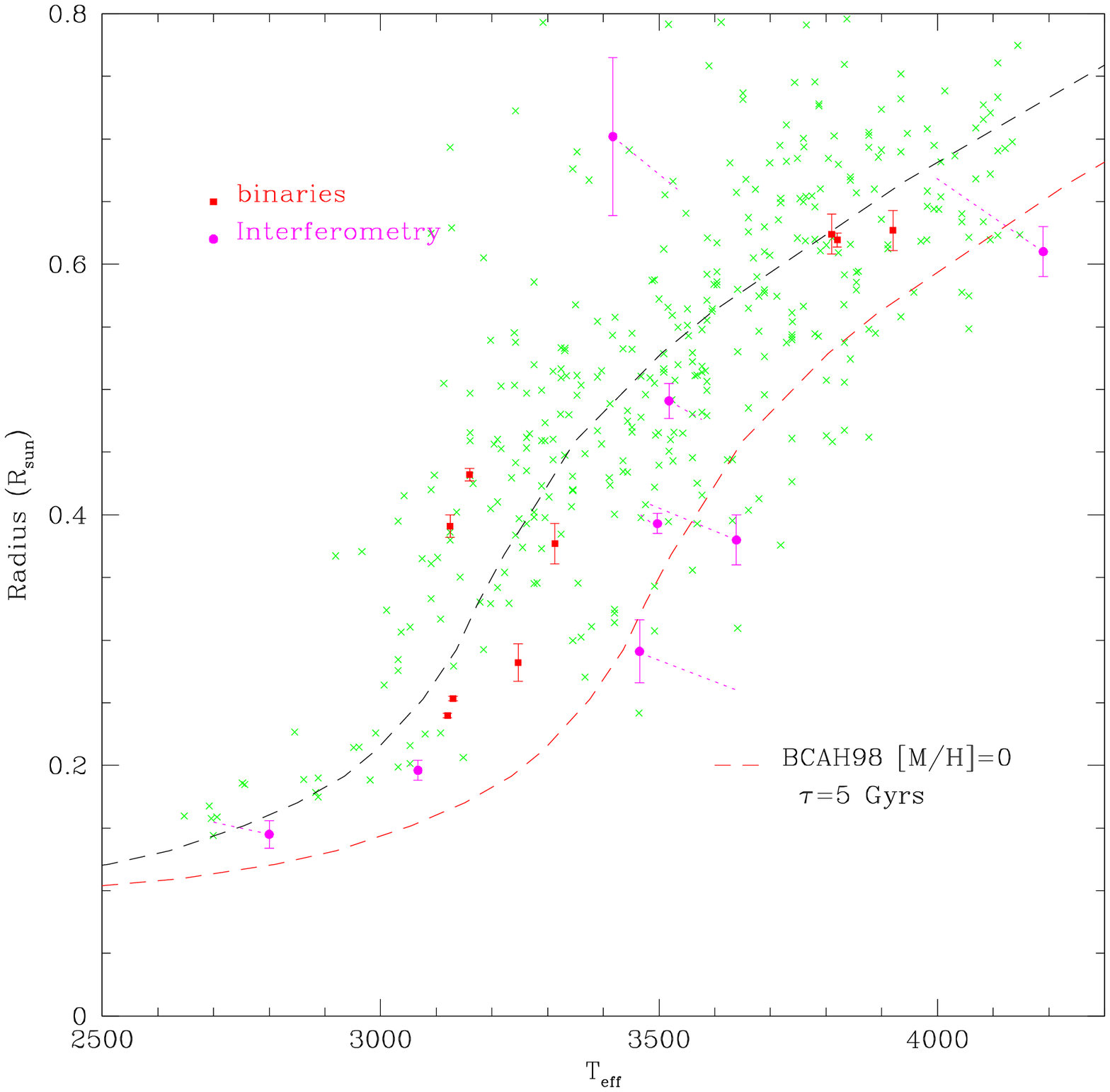}
\caption{\label{fig:radius-lum}
{\it Left:} Radius-luminosity relation for M dwarfs. 
Crosses are nearby field M dwarfs with reliable 
parallaxes, and temperatures derived from $V-I$ colors.
The relationship is smooth across the M spectral types, as predicted by 
stellar evolution models (dashed lines). 
{\it Right:} Radius-$T_{\rm eff}$ relation. As highlighted
\citet{Clemens1998}, there is a marked decrease in radius near
spectral type M4. Note, however, that this change in the 
radius-$T_{\rm eff}$ relationship is not infinitely sharp, but rather is a
finite transition from spectral types M3 to M6.
}
\end{figure}

Neill Reid discussed the observed properties of M dwarfs across the
fully convective boundary.
It is useful to recall up front that certain fundamental stellar 
properties do not evince any obvious manifestation of the changes in
stellar structure that accompany the transition into the fully convective
regime. For example, the mass-luminosity relation is smooth through the
M spectral types. Similarly, Figure~\ref{fig:radius-lum}a shows the 
radius-luminosity relation for a large sample of M dwarfs 
\citep{Ribas2006,Demory2009}: The 
observed radius-luminosity relation is smooth through the M spectral types,
as predicted by stellar evolution models. Evidently, energy generation 
and output in these stars does
not much know or care about changes in structure or in energy transport
that may be occurring across the fully convective boundary.

However, other basic stellar properties do manifest clear changes across
a mass of $\sim$0.3 M$_\odot$. Figure~\ref{fig:radius-lum}b shows the
radius-temperature relation for the same stars in Figure~\ref{fig:radius-lum}a.
Unlike the featureless radius-luminosity relation, here
we see a clear change in slope---with the stellar radii decreasing rapidly
with decreasing effective temperature---centered around 
$T_{\rm eff} \sim 3150$~K (spectral type M4).  The change in stellar 
radius here is a factor of $\sim$2.  In other words, there is a relatively
large range of stellar mass and radius over a relatively small range of 
spectral types.  Indeed, 
stars with a wide range of masses around the fully convective transition
are predicted to have similar spectral types 
\citep[cf.\ Fig.\ 4 in][]{chabrier2000}, a consequence of the 
flatness of the mass-temperature relation caused by H$_2$ formation.
There is a strong morphological similarity between the observed and predicted 
relations, but the two are offset by $\sim$250K, with the observations
cooler than the models (compare pink and black curves in
Fig.~\ref{fig:radius-lum}).

Notably, the change in the radius-temperature relation near M4 
is not an abrupt step function; rather, it is a {\it finite transition}
between $T_{\rm eff} \approx 3300$~K and $T_{\rm eff} \approx 2850$~K 
(i.e.\ spectral types M3 to M6).
At later spectral types, the stellar radii asymptotically approach the
value of $\approx$0.1 R$_\odot$ dictated by electron degneracy pressure.

\begin{figure}[ht]
\plottwo{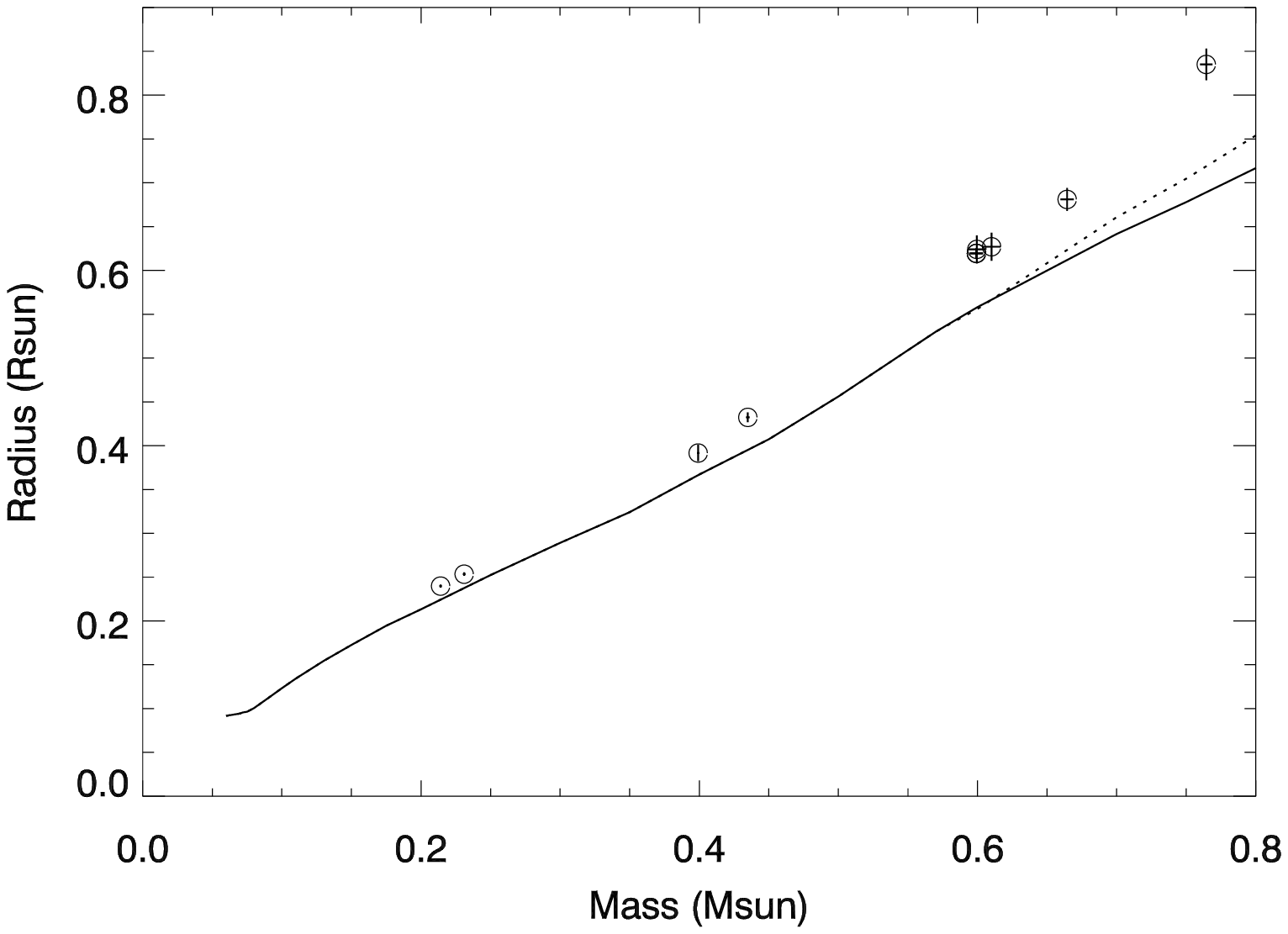}{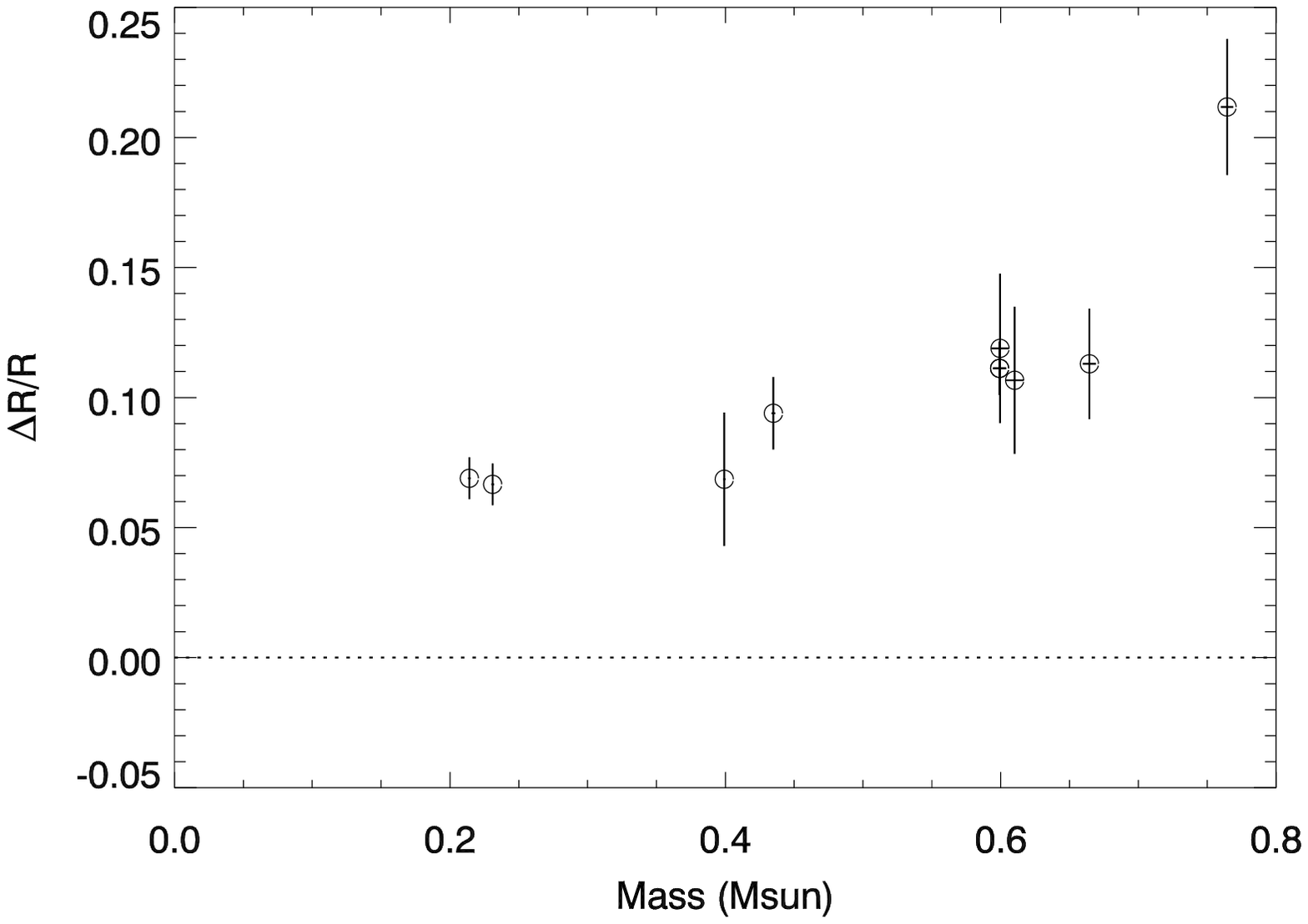}
\caption{\label{fig:leslie}
{\it Left:} Mass-radius relation for eclipsing binary systems
with masses and radii measured to better than 3\%.
{\it Right:} 
Fractional increase in the radii over the theoretically predicted radii
\citep{Baraffe1998}.
The enlarged radii are thought to be caused by the affects of magnetic fields.  
Early type M~dwarfs show a larger increase in radii compared to late-type,
fully convective M~dwarfs, consistent with predictions of theoretical models
\citep{Mullan2001,Chabrier2007}. 
}
\end{figure}

As Leslie Hebb discussed, recent
direct radius measurements of M dwarfs in eclipsing binaries
show the stellar radii to be inflated relative to the radii predicted by
theoretical stellar evolution models \citep[e.g.][]{Ribas2006}. 
This has been attributed to
the effects of stellar magnetic fields and activity \citep[e.g.][]{Morales2008}.
The magnitude of this effect does change across the fully convective 
boundary (Figure~\ref{fig:leslie}), but as above it does not appear to 
change in an abrupt fashion.

\subsection{Stellar activity and stellar rotation: Effects of stellar age}
As Jonathan Irwin and Richard Jackson discussed, recent observational
results have confirmed that the M4 transition plays an important part
in shaping the rotation and magnetic activity evolution of M dwarfs. 
Large samples of M dwarfs show an increase in activity fractions 
around spectral types of $\sim$M4 \citep{West2008}.

\begin{figure}[ht]
\plotfiddle{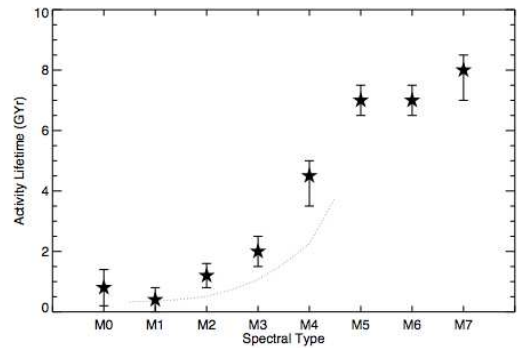}{1.5in}{0}{43}{43}{-100}{-8}
\caption{\label{fig:west}
Activity lifetime as a function of spectral type for M dwarfs 
(West et al.\ 2008). The lifetime over which an M dwarf stays active is a
strong function of spectral type, from 1--2 Gyr for M3 dwarfs to 7--8 Gyr
for M6 dwarfs. Evidently, activity lifetime does respond to the change in
internal structure across the fully convective boundary, but the transition
is a finite one between M3 and M6.
}
\end{figure}

This result has been shown to be dependent on the ages of the stars. When 
looking at the activity fractions as a function of the M dwarfs' location 
above the Galactic plane, and using a simple model of dynamical heating 
which scatters M dwarfs to larger Galactic scale heights over time, 
\citet{West2008} determine the amount of time that M dwarfs of
different spectral types remain chromospherically active. They find that the 
``activity lifetime" changes dramatically between spectral types M3 and M6 
(from 1--2 Gyr for M3 dwarfs to 7--8 Gyr for M6 dwarfs; Fig.\ref{fig:west}).

\begin{figure}[ht]
\plotfiddle{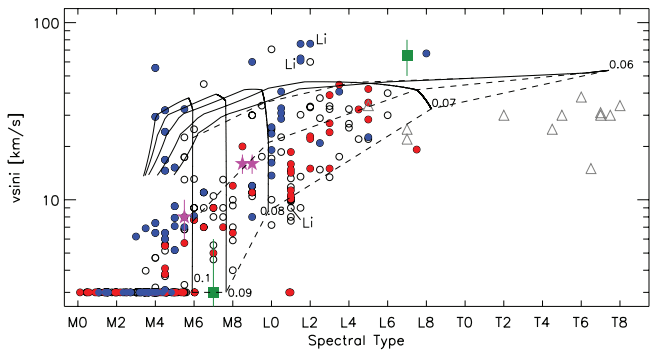}{1.5in}{0}{40}{40}{-135}{-10}
\caption{\label{fig:reiners}
Rotation ($v\sin i$) of late-type stars \citep{Reiners2009}.
Starting at spectral type $\sim$M3, mean $v\sin i$ increases monotonically,
implying that the spindown timescale increases monotonically across the
fully convective boundary.
}
\end{figure}

This important
transition was bolstered by the rotation analysis of \citet{Reiners2009},
who found that the average rotation velocities of a nearby sample of M dwarfs 
increases dramatically and monotonically as one looks at later spectral types.
The average $v\sin i$ is $<$3 km/s at M2, increases to $\sim$10 km/s at M6,
and continues increasing to $\sim$30 km/s at L4 (Figure~\ref{fig:reiners}).
This almost certainly points to a lengthening of the timescale for
stellar spindown as the stellar structure changes from solar-type (i.e.\
radiative core $+$ convective envelope) to fully convective.
However, the transition is not a step function, thus again pointing to the
importance of evolution over time.

\begin{figure}[ht]
\plotfiddle{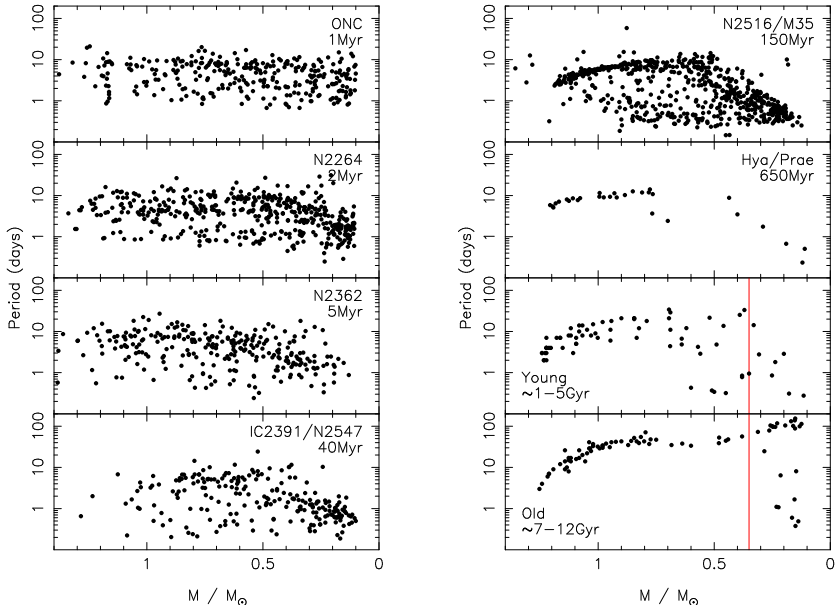}{2.8in}{-90.}{45.}{45.}{-175}{250}
\caption{\label{fig:irwin}
Observed rotation period distributions for low-mass stars in clusters
of varying ages. Among clusters with ages $>$100 Myr, two sequences of rapid 
and slow rotators are observed, with the slow-rotator branch becoming 
increasingly dominant with time. At the fully convective boundary (vertical 
line), the stars do not transition to the slow-rotator sequence
instantaneously, but rather on a finite timescale.
}
\end{figure}

The importance of time evolution is seen most clearly in rotation-period
distributions in populations of different ages.
Fig.~\ref{fig:irwin} shows the evolution of stellar rotation periods
from $\sim$1 Myr to $\gtrsim$10 Gyr. 
The situation among the pre--main-sequence populations is complex, likely
due to the effects of multiple competing phenomena \citep{bouvier}. However, 
at ages of $\sim$100 Myr and older, the stars show a clear pattern with
two dominant sequences of rapid and slow rotators; the slow-rotator
sequence becomes increasingly dominant with increasing stellar age. 
These are the so-called `C' and `I' sequences of the Barnes 
``gyrochronology" paradigm \citep{Barnes2003,Barnes2007}.

Importantly, it is also clear from this figure that at all ages there is 
a spread of stellar rotation periods near the fully convective boundary;
age alone does not determine whether a given M4 star will be on the `C'
sequence or on the `I' sequence. Thus, while all stars appear to follow
a similar pattern of evolution in stellar rotation, presumably connected to 
evolution of their interior structure and magnetic field properties, there 
is evidently star-to-star scatter in these properties at any given age.

\subsection{Magnetic fields: Topology and effects on stellar properties}
\begin{figure}[ht]
\plottwo{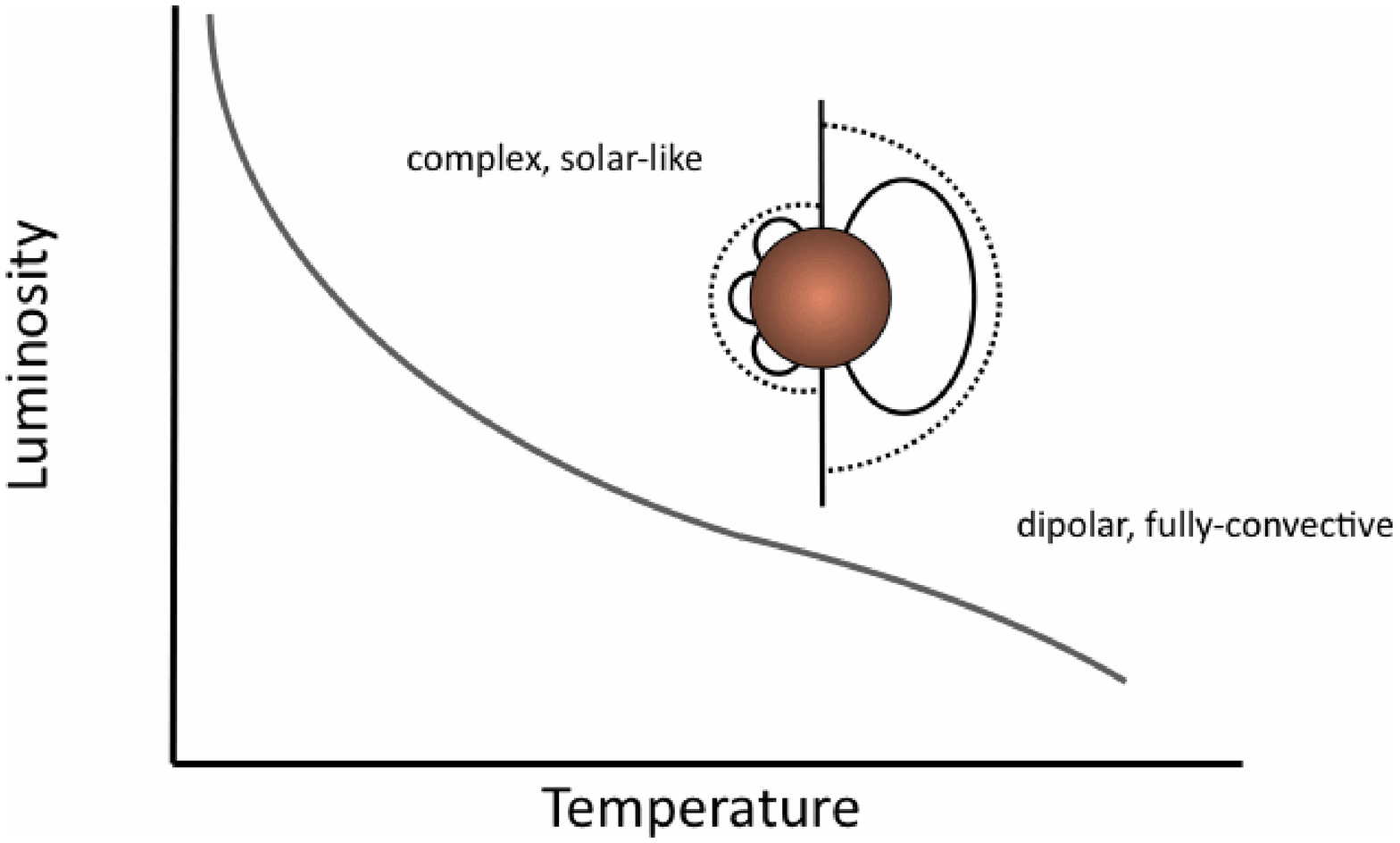}{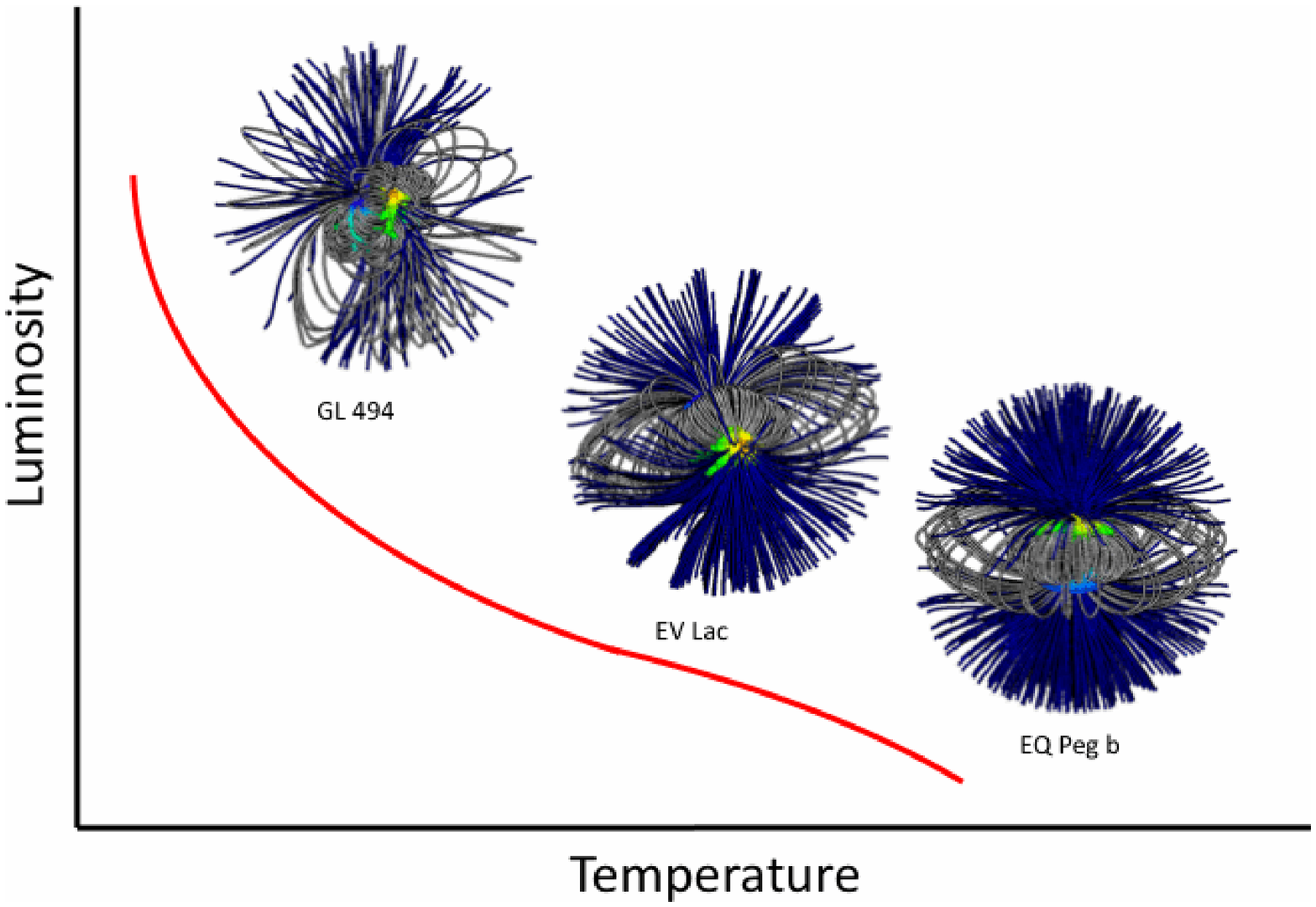}
\caption{\label{fig:moira-cartoon}
{\it Left:} Cartoon representing observed differences in surface magnetic
field topology across the fully convective regime. 
{\it Right:} Reconstructed surface field topologies for three low-mass 
stars are shown (top to bottom: Gl 494, EV Lac, EQ Peg b).
The field changes from a strong, highly
structured, multi-polar configuration at early M~types, to a weaker, more
well organized dipole at the late M~types.
}
\end{figure}

As discussed by Moira Jardine and Julien Morin, M~dwarfs also show clear
topological changes in their surface magnetic fields as a function of
spectral type.
Figure~\ref{fig:moira-cartoon}a shows a simple cartoon illustration
summarizing the basic change that is observed: 
from globally weak but highly structured (i.e.\ multipolar 
and non-axisymmetric) fields at 
early spectral types, to globally strong and well ordered
(i.e.\ dipolar) fields at later spectral types.
In other words, the energy in the dipole component, relative to higher-order
components, is higher in the lower-mass stars.
Figure~\ref{fig:moira-cartoon}b depicts these changes in reference to three
specific low-mass stars (Gl~494, EV Lac, EQ Peg b) whose magnetic field 
topologies have been modeled in detail \citep{jardine2002} using Zeeman 
Doppler Imaging (ZDI) maps as boundary conditions \citep{Donati2008,Morin2008}.

\begin{figure}[!ht]
\plotfiddle{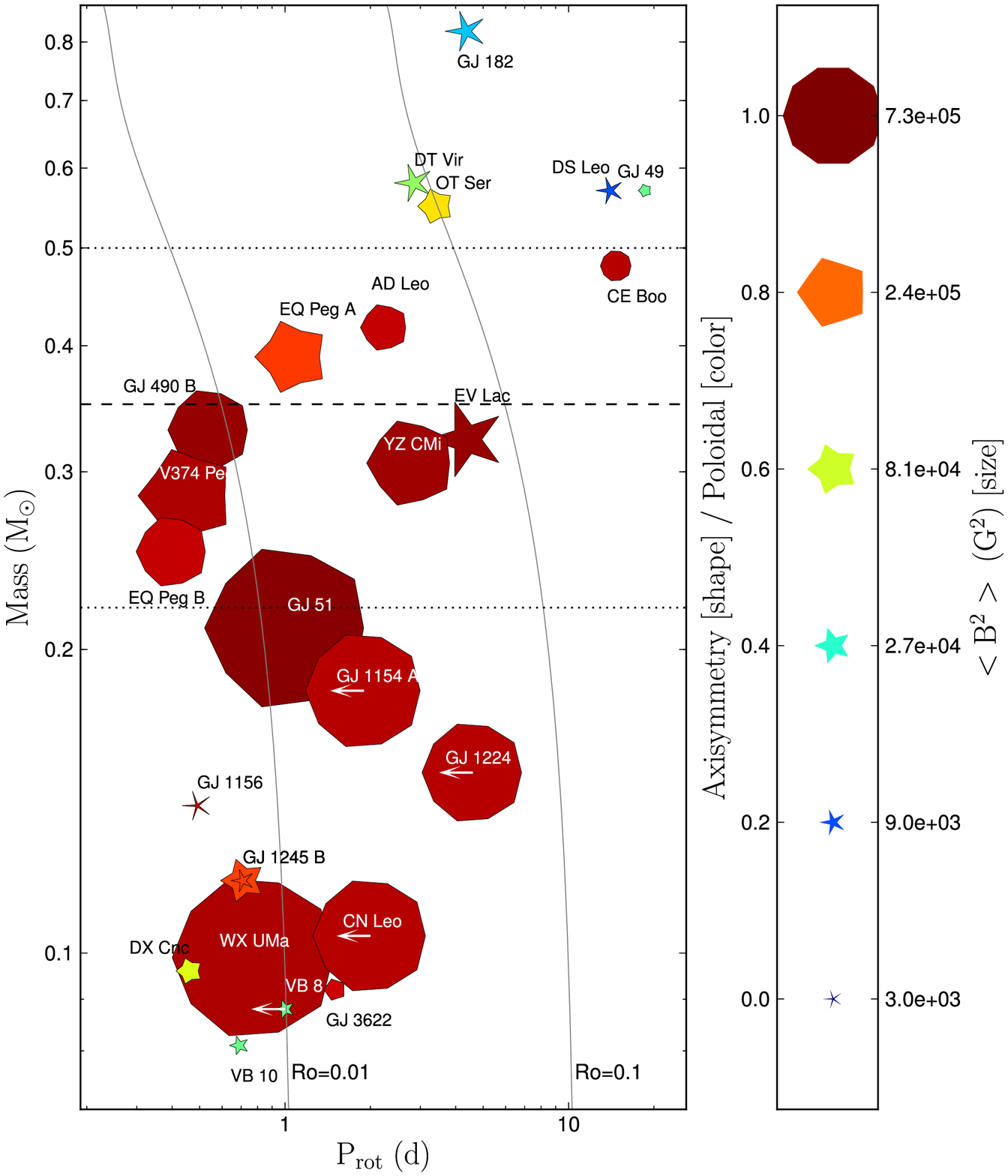}{4.3in}{0.}{50}{50}{-150}{-43}
\caption{\label{fig:morin}
Properties of the large-scale
magnetic topologies of a sample of M dwarfs inferred  
from spectropolarimetric observations and Zeeman Doppler Imaging (ZDI)
\citep{Donati2008,Morin2008,Phanbao2009} as a 
function of rotation period and mass. Larger symbols indicate stronger  
fields, symbol shapes depict the degree of axisymmetry of the  
reconstructed magnetic field (from
decagons for purely axisymmetric to sharp stars for purely non  
axisymmetric), and colours the field configuration (from blue for  
purely toroidal to red for purely poloidal).  
Leftward white arrows indicate stars for which only an upper limit on 
$P_{\rm rot}$ is derived and the field properties are extrapolated. 
For VB8 the same symbol as VB10 is arbitrarily used, no magnetic field 
was detected from spectropolarimetric observations. For GJ 1245 B symbols  
corresponding to 2007 and 2008 data are superimposed in order to  
emphasize the variability of this object. Solid lines represent  
contours of constant Rossby number $R_0 = 0.1$ (saturation threshold) and  
0.01. The theoretical full-convection limit (M$\sim 0.35$ M$_{\odot}$,
\citet{Chabrier1997}) is plotted as a horizontal dashed line, and  
the approximate limits of the three stellar groups identified are  
represented as horizontal dotted lines.
}
\end{figure}

This is shown more quantitatively in Figure~\ref{fig:morin}, which
depicts the field topology and strength as a function of stellar mass,
derived for a large number of low mass stars using the ZDI technique.
Interestingly, we see a hint of two transitions here. The first occurs near
the fully convective boundary, where as already mentioned the fields
transition from a globally weak, highly structured topology, to a globally
strong, dipolar topology.
A second transition appears at spectral type M6--M8, such that
below M$\sim 0.15$~M$_{\odot}$, stars with similar stellar masses and
rotational properties can have very different magnetic topologies. 
This apparent M6--M8
transition could be reflecting the point at which the stellar atmospheres 
become completely neutral \citep[e.g.][]{Mohanty2003}, though it could
again reflect an age effect as witnessed in the bifurcated rotational
properties of M dwarfs (see Fig.~\ref{fig:irwin}).

Finally, it is important to bear in mind that the ZDI measurements,
because they rely upon observations of polarized light, are probing 
field components that represent only a few percent of the total magnetic
energy. There is likely a large amount of magnetic flux in small-scale
field components (which do not contribute much net polarized light) 
in addition to the larger scale components that the ZDI technique is 
mapping.

Dermott Mullan discussed recent modeling efforts to account for the
observed inflated radii of M dwarfs compared to the predictions of most
(non-magnetic) theoretical stellar evolution models.
Chabrier and collaborators have suggested that strong {\it surface} fields 
can inhibit surface convection and also create cool starspots, 
both of which act to
decrease the surface temperature and then---because the stellar luminosity
generated in the core is unaffected---requires an increase in the stellar
radius.

\begin{figure}[ht]
\plotfiddle{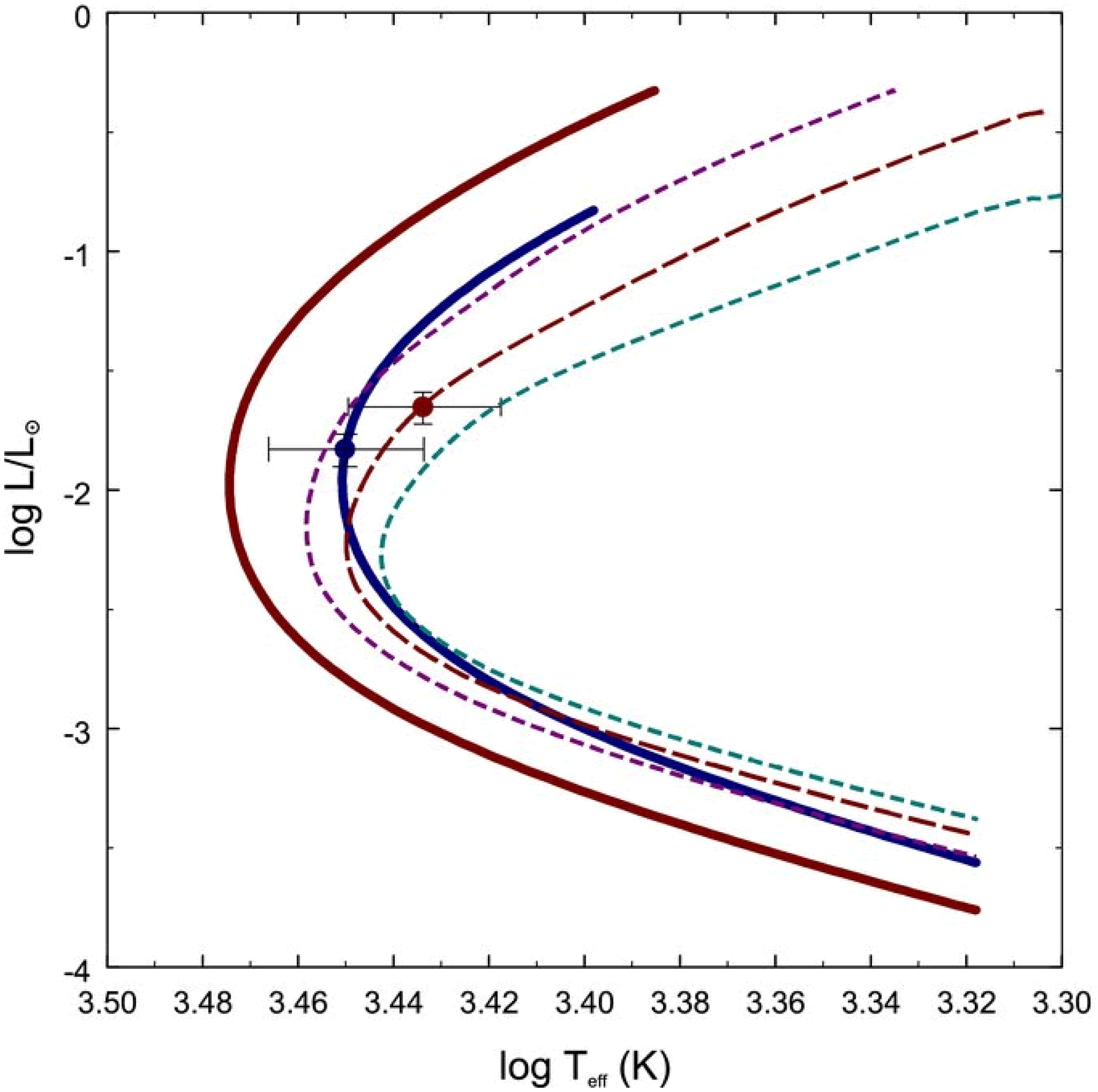}{2.5in}{0.}{35.}{35.}{-110}{-40}
\caption{\label{fig:bdeclipser}
The reversal of effective temperature with mass for the brown-dwarf 
eclipsing binary 2M0535--05 \citep{Stassun2006,Stassun2007} can be explained
by recent models that include the effects of strong magnetic fields threading
through the entire interior of a low-mass star \citep{MacDonald2009}.
Solid curves show the non-magnetic models for the primary (more massive)
brown dwarf (red) and the secondary brown dwarf (blue), dashed lines show 
the magnetic models.
}
\end{figure}

\begin{figure}[ht]
\plotone{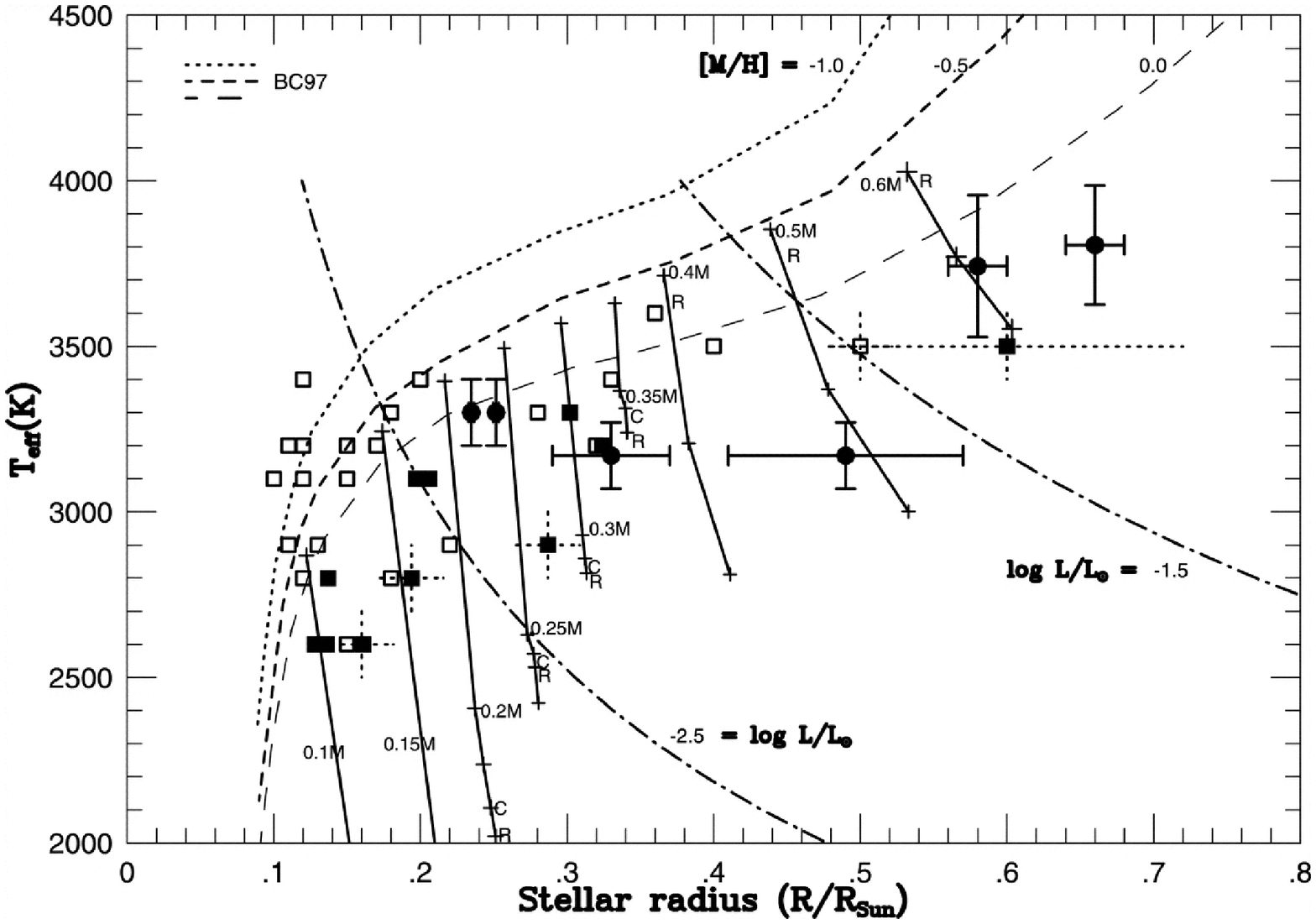}
\caption{\label{fig:mullan}
Models of \citet{Mullan2001} showing the change in stellar
radius, temperature, and luminosity, resulting from strong internal magnetic
fields.
}
\end{figure}

Mullan also discussed alternate models in which a strong field threads the
entire star. These models can successfully reproduce for example the observed
temperature reversal in the brown-dwarf eclipsing binary 2M0535--05
(Figure~\ref{fig:bdeclipser}), which at an age of only $\sim$1 Myr has a
spectral type of M6.
These models do predict a change in the stellar luminosity as a result of
the very strong ($\sim$MG) fields in the core (Figure~\ref{fig:mullan}).

\section{Putting it all together: A conceptual framework}
\begin{figure}[!ht]
\plotone{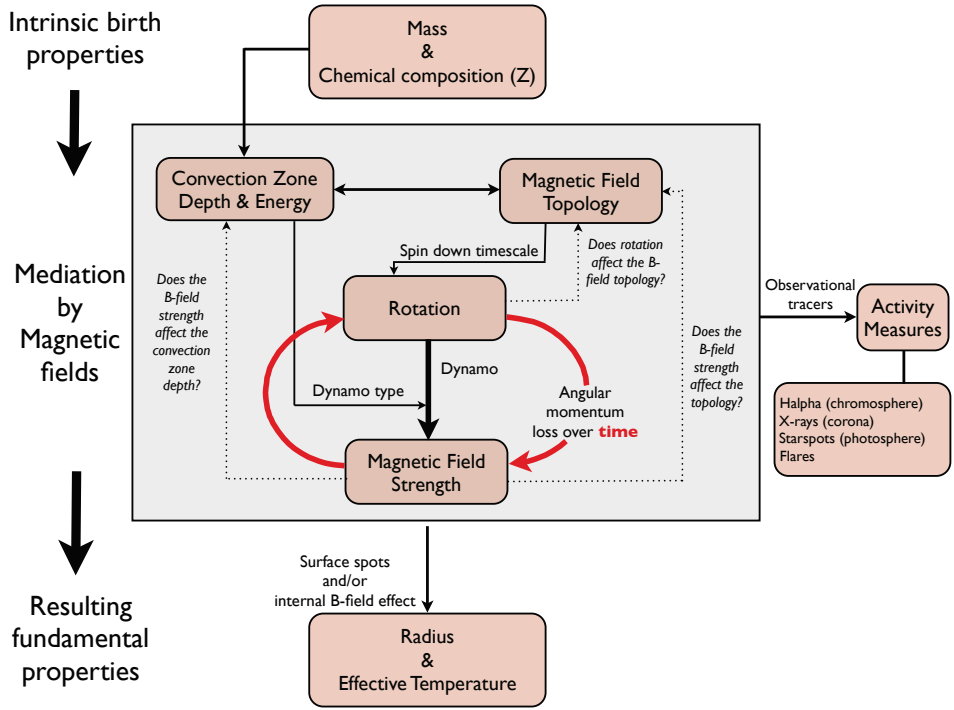}
\caption{\label{fig:flowchart}
Representation of the process by which magnetic fields influence
the interior structure and observed fundamental properties of M~dwarf stars.
M~dwarfs are defined intrinsically by mass and chemical composition which
determine the initial depth of the convection zone.  This in turn affects
the magnetic field topology and strength through the dynamo type 
($\alpha$--$\Omega$ or $\alpha^2$).  The rotation rate, which is moderated 
by angular momentum loss through stellar winds determined largely by
the field topology, in turn affects the field strength for stars in
the non-saturated regime via the dynamo process.  X-rays, chromospheric
emission, starspots, and flaring all trace the magnetic activity.  Over time,
stars spin down and the magnetic field weakens, causing the entire system
to evolve.  It remains unclear how much feedback there is in the system.  
Does the
strength of the magnetic field affect the topology or the convection zone
depth, and how much does the rotation rate influence the field topology?
Taken together, the interlinked magnetic fields, internal structure, and
rotation properties affect the resulting fundamental parameters: radius
and temperature.  However, the detailed mechanism by which the latter
occurs is not yet fully understood.
}
\end{figure}

Figure~\ref{fig:flowchart} is
an initial attempt to conceptually summarize the inter-related 
phenomena discussed during our Session. This is not intended to be
inclusive of all mechanisms or observable phenomena possibly
relevant to M dwarfs, and it is by necessity a simplified representation of 
what is likely to be a more complex set of inter-relationships. 

The figure attempts to represent the complex set of causal 
relationships---stellar structure, magnetic field topology, rotation, 
magnetic field strength---that mediate the fundamental relationships
between basic stellar properties (e.g., how mass and composition 
ultimately translate into radius and effective temperature).
As depicted in the figure, we imagine that, absent any magnetic field
effects, there would be a single, fundamental mass--radius--temperature
relationship (modulo metallicity). This relationship is, however, 
modified by the presence of 
a magnetic field that manifests itself on the surface through various
activity tracers (e.g.\ chromospheric H$\alpha$ emission). This magnetic
field affects the observed surface properties of the star (effective
temperature and radius) via surface effects (i.e.\ star spots;
\citet{Chabrier2007}) and/or via a strong field that inhibits convection
throughout the stellar interior \citep{Mullan2001}.

The mechanism that drives the magnetic field's action (and
perhaps its generation) is
likely complex, but almost certainly involves at its heart a rotational 
dynamo. We envision that the star's rotation directly affects the strength
of the magnetic field (more rapid rotation $\rightarrow$ stronger field)
for stars below the saturation threshold.
In general, a stronger field will drive a more rapid evolution of the star's
angular momentum, leading to a decline of the star's rotation on a certain
spindown timescale, and this spindown in turn feeds back into the strength 
of the field, dialing it down over time.

Importantly, the magnetic field acts according to an intrictate feedback 
process such that the magnetic field impacts the stellar structure 
(i.e.\ convection zone depth) and rotation
rate of the star which in turn affects the (re)generation of the field.
The nature of the dynamo
itself depends on the interior structure (i.e., convection zone depth).
As the star approaches a more
fully convective state (that is, as the size of the radiative zone
recedes), there must be a transition from a solar-type ($\alpha-\omega$)
dominated dynamo to a purely turbulent ($\alpha^2$) dynamo.
The spindown timescale is probably intimately coupled to the
field topology, but exactly how the two are connected is not obvious. 
For example,
later type M~dwarfs have longer spindown timescales, yet these same 
stars have magnetic field topologies which posses
more open field lines (i.e.\ dipolar fields), which would seem to enable
them to more effectively drive mass and angular momentum loss.
Clearly, we do not yet understand entirely how magnetic topology
affects the observed rotational evolution of these stars.

It is moreover not clear whether the dynamo-driven strength of the field can 
feed back into the
depth of the convection zone and/or the field topology. For example, 
it is thought that a very strong field can cause an otherwise fully
convective star to generate a radiative core, either directly through the 
inhibition of convection at the center of the star \citep[e.g.][]{Mullan2001}
and/or indirectly through altering the boundary conditions at the stellar 
surface \citep[e.g.][]{Chabrier2007}. Similarly, in some models very rapid 
stellar rotation can cause the opening of surface field lines 
\citep[e.g.][]{Jardine2010}.
Thus, the inter-relationships between rotation, field strength,
interior structure, and surface field topology are likely complex, non-linear,
and in any event dependent on the age of the star through its rotational
history.

The rotation-period distributions of low-mass stars as a function of stellar
age show that rotation is not simply a function of time alone. Indeed, current
thinking identifies at least two dominant sequences of rotational properties
in clusters (e.g., the `C' and `I' sequences of Barnes), such that a star of
a given mass and age is likely to have one of two possible rotation periods.
Perhaps these distinct rotational sequences reflect stars in different states
of the complex feedback mechanisms between field generation, dynamo type, 
and field topology explored above. Indeed, the ZDI maps of field topologies
appear to bear out this picture of multiple types of field configurations
intermixed among the M3--M6 spectral types (Figure~\ref{fig:morin}). 
Given this, and in light of the time-dependence of these effects, it is
perhaps not surprising that the ``M4 transition" across the fully convective
boundary is not a sharp transition, but rather ``smeared" from approximately
M3 to M6. For example, the two eclipsing binaries CM Dra and CU Cnc
(see Fig.~\ref{fig:radius-lum}),
with masses and radii that differ by a factor of $\sim$2 but with nearly 
identical effective temperatures (spectral types $\approx$M4), may be 
exemplars of this smearing effect.

\section{Open questions and future work}

We concluded with an open discussion leading to a set of
questions to help guide ongoing work by the community interested in the
``M4 transition" and related phenomena.

\begin{enumerate}
\item What field strengths are necessary and reasonable in the interiors 
of stars to potentially explain the observed mass-radius relationship? 
\item Abundance measurements needed! (e.g.\ [O/Fe])  
The mass-radius relationship is actually a mass-radius-abundance relationship.
FGK$+$M binaries are a promising approach, tying M dwarf abundances to 
FGK primary stars.
\item Continue to identify low-mass eclipsing binaries. 
Long-period systems are needed to disentangle effects of 
magnetic activity on stellar radii and temperatures.
\item Examine selection effects in samples, especially with regard to activity.
Ages of stellar samples matters.
\item What is the mass of the fully convective boundary?
Depends on rotation/activity/age? 
\item Other observational clues: 
Period gap in CVs (2--3 hr periods) $\rightarrow$ kink in mass-radius 
relationship?
\item Independent radius measurements: interferometry, 
$P_{\rm rot} + v\sin i$ (gives $R \sin i$)
\item Bulk of field is in very small-scale structures $\rightarrow$ 
How to fully characterize the fields? 
Constraints from many techniques/tracers are needed to fully characterize 
magnetic fields: Zeeman broadening, spectropolarimetry, radio, X-rays, etc.
\item ``M6--M8 transition" in field topologies, caused by
neutral atmospheres? Age?
\item 
How does the fully- to partly-convective evolution on the 
pre--main-sequence affect
the early magnetic and rotational history of low-mass stars?
\end{enumerate}

\bibliography{stassun_k}

\end{document}